\documentclass[showpacs,prl,floatfix,twocolumn]{revtex4}
\usepackage{graphicx,amssymb,amsmath}

\begin{document}

\title{Scaling in transportation networks}

\author{R\'emi Louf}
\email{remi.louf@cea.fr}
\affiliation{Institut de Physique Th\'{e}orique, CEA, CNRS-URA 2306, F-91191, Gif-sur-Yvette, France}

\author{Camille Roth}
\affiliation{Centre Marc Bloch Berlin (An-Institut der Humboldt Universit\"at, UMIFRE CNRS-MAE), Friedrichstr 191, D-10117 Berlin}
\affiliation{Centre d'Analyse et de Math\'ematique Sociales, EHESS-CNRS (UMR 8557),190-198 avenue de France, FR-75013 Paris, France}

\author{Marc Barthelemy}
\email{marc.barthelemy@cea.fr}
\affiliation{Institut de Physique Th\'{e}orique, CEA, CNRS-URA 2306, F-91191, Gif-sur-Yvette, France}
\affiliation{Centre d'Analyse et de Math\'ematique Sociales, EHESS-CNRS (UMR 8557),190-198 avenue de France, FR-75013 Paris, France}

\begin{abstract}
Subway systems span most large cities, and railway networks most countries in the world. These networks are fundamental in the development of countries and their cities, and it is therefore crucial to understand their formation and evolution. However, if the topological properties of these networks are fairly well understood, how they relate to population and socio-economical properties remains an open question. We propose here a general coarse-grained approach, based on a cost-benefit analysis that accounts for the scaling properties of the main quantities characterizing these systems (the number of stations, the total length, and the ridership) with the substrate's population, area and wealth. 
More precisely, we show that the length, number of stations and ridership of subways and rail networks can be estimated knowing the area, population and wealth of the underlying region. These predictions are in good agreement with data gathered for about $140$ subway systems and more than $50$ railway networks in the world. We also show that train networks and subway systems can be described within the same framework, but with a fundamental difference: while the interstation distance seems to be constant and determined by the typical walking distance for subways, the interstation distance for railways scales with the number of stations. 
\end{abstract}

\maketitle

\section*{Introduction}

Almost $200$ subway systems run through the largest agglomerations in the world and offer an efficient alternative to congested road networks in urban areas. Previous studies have explored the topological and geometrical static properties of these transit systems~\cite{Benguigui:1992,Benguigui:1995,Derrible:2009,Sienkiewicz:2005,Levinson:2012}, as well as their evolution in time~\cite{VonFerbe:2009,Roth:2012,Leng:2014}. However, subways are not mere geometrical structures growing in empty space: they are usually embedded in large, highly congested urban areas and it seems plausible that some properties of these systems find their origin in the interaction with the city they are in. Previous studies~\cite{Levinson:2008,Xie:2009} have indeed shown that the growth and properties of transportation networks are tightly linked to the characteristics of urban environment. Levinson \cite{Levinson:2008} for instance, showed that rail development in London followed a logic of both `induced supply' and `induced demand'. In other words, while the development of rail systems within cities answers a need for transportation between different areas, this development also has an impact on the organisation of the city. Therefore, while the growth of transportation systems cannot be understood without considering the underlying city, the development of the city cannot be understood without considering the transportation networks that run through it. As a result, the subway system and the city can be thought as two systems exhibiting a symbiotic behaviour. Understanding this behaviour is crucial if we want to gain deeper insights into the growth of cities and how the mobility patterns organise themselves in urban environments.

At a different scale, railway networks answer a need for fast transportation between different urban centers, and we therefore expect their properties to be linked to the characteristics of the underlying country. A model of growth has been recently proposed~\cite{Louf:2013}, and relates the existence of a given line to the economical and geographical features of the environment. An interesting question is thus to know whether subways and railway networks behave in the same way, but at different scales. In other words, we are interested to know whether subways are merely scaled down railway networks, or whether they are fundamentally different objects, following different growth mechanisms. Also, the existence of scaling between the system's output and its size is important as it suggests that very general processes are governing the growth of these networks \cite{Banavar:1999,Louf:2014}.

Although many studies~\cite{Kansky:1963,Derrible:2009,Levinson:2012} explore the interplay between regional characteristics and the structure of transportation networks, a simple picture relating the network's most basic quantities and the region's properties is still lacking. In the spirit of what has recently been done for cities~\cite{Louf:2014} and for railway networks \cite{Black:1971,Louf:2013}, we propose here a large-scale framework and try to understand how subways and railway networks scale with some of the substrates' most basic attributes: population, surface area and wealth. As a result, we are able to relate the total ridership, the number of stations, the length of the network to socio-economical features of the environment. We find that these relations are in good agreement with the data gathered for $138$ subway systems and $58$ railway networks accross the world. In particular, we show that even if the main mechanisms are the same, the fact that both systems operate at different scales is responsible for their different behaviors. We believe this should lay the foundations for more specific and involved discussions.

\section*{Results}

\subsection*{Framework}

A transportation network is at least characterized by its total number of nodes (which are here train or subway stations), its total length, and the total (yearly) ridership. On the other hand, a city (or a country in the railway case) is characterized by its area, its population and its Gross Domestic Product (GDP). Because transportation systems do not grow in empty space, but result from multiple interactions with the substrate, an important question is how network characteristics and socio-economical indicators relate to one another. Naturally, a cost-benefit analysis seems to be the appropriate theoretical framework. This approach has been developed in the context of the growth of railway networks~\cite{Black:1971,Louf:2013}, and in these studies an iterative growth was considered: at each step an edge $e$ is built such that the cost function
\begin{equation}
Z_e = B_e - C_e
\end{equation}
is maximum. The quantity $B_e$ is the expected benefit and $C_e$ the expected cost of edge $e$. In the following, we consider networks after they have been built, and we assume that they are in a `steady-state' for which we can write a cost function of the form
\begin{equation}
Z =\sum_eZ_e= B - C
\end{equation}
where $B$ is the total expected benefits and $C$ the total expected costs, mainly due to maintenance (in the steady state regime). We further assume that, during this steady-state, operating costs are balanced by benefits. In other words
\begin{equation}
Z \approx 0
\end{equation}
Indeed, because lines and stations cost money to be maintained, we expect the network to adapt to the way it is being used. Therefore we can reasonably expect that at first order the cost of operating the system is compensated by the benefits gained from its use. In the following we will apply this general framework to subway and railway networks in order to determine the behavior of various quantities with respect to population and GDP.

\subsection*{Subways}

In the case of subways, the total benefits in the steady-state are simply connected to the total ridership $R$ and the ticket price $f$ over a given period of time. The costs, on the other hand, are due to the maintenance costs of the lines and stations, so that we can write (for a given period of time)
\begin{equation}
Z_{sub} = R\,f - \epsilon_L L - \epsilon_S N_s
\label{eq:cost-benefit}
\end{equation}

where $L$ is the total length of the network, $\epsilon_L$ the maintenance cost of a line per unit of length, $N_S$ the total number of stations and $\epsilon_S$ the maintenance cost of a station (for a given period time).

It is usually difficult to estimate the ridership of a system given its characteristics and those of the underlying city. Due to the importance of such estimates for planning purposes, the problem of estimating the number of boardings per station given the properties of the area surrounding the stations has been the subject of numerous studies~\cite{Matsunaka:2013,Kuby:2004}. Here we are interested in the dependence of global, average behavior of the ridership on the network and the underlying city. Very generally, we write that the number $R_i$ of people using the station $i$ will be a function of the area $C_i$ serviced by this station---the `coverage'~\cite{Derrible:2009}---and of the population density $\rho = \frac{P}{A}$ in the city
\begin{equation}
R_i = \xi_i\, C_i\, \rho
\end{equation}
where $\xi_i$ is a random number of order one representing the fraction of people that are in the area serviced by the station and who use the subway. The main difficulty is in finding the expression of the coverage. It depends, a priori, on local particularities such as the accessibility of the station, and should thus vary from one station to another. We take here a simple approach and assume that on average
\begin{equation}
C_i \sim \pi\, d_0^{\,2}
\end{equation}
where  $d_0$ is the typical size of the attraction basin of a given station. If we assume that it is constant, the total ridership can be written as
\begin{equation}
R = \sum_i R_i \sim \overline{\xi} \pi d_0^2 \rho \: N_s
\label{eq:ridership}
\end{equation}
where $\overline{\xi} = \frac{1}{N_s}\,\sum_i \xi_i$ is of the order of 1.\\

\begin{figure*}
\centering
\includegraphics[height=150px]{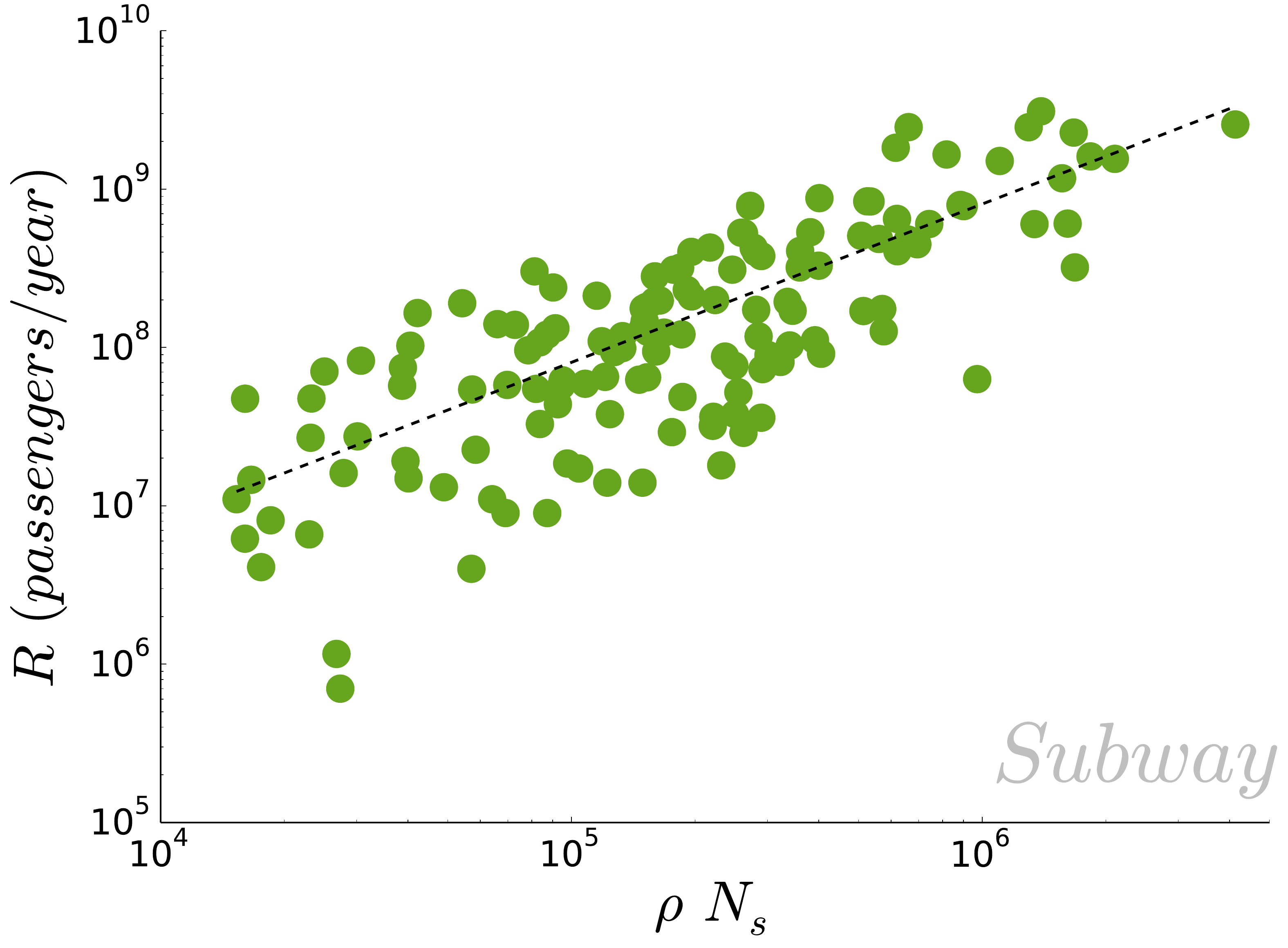}
\includegraphics[height=150px]{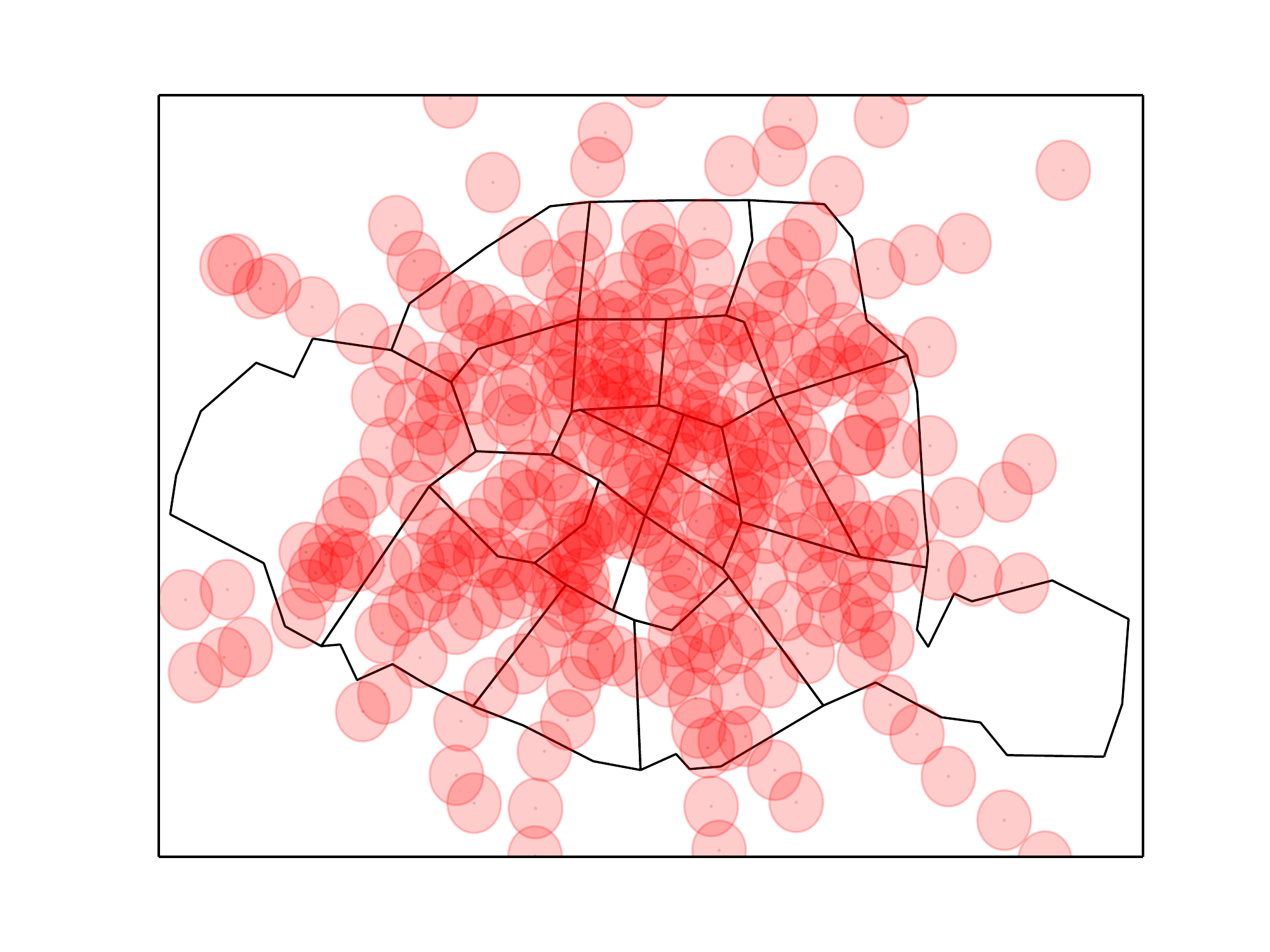}
\caption{{\bf (Subway) Relationship between ridership and coverage} (Left) We plot the total yearly ridership $R$ as a function of $\rho\,N_s$. A linear fit on the $138$ data points gives $R \approx 800\:\rho N_s$ ($R^2=0.76$) which leads to a typical effective length of attraction $d_0 \approx 500\,\text{m}$ per station. (Right) Map of Paris (France) with each subway station represented by a red circle of radius $500\,\text{m}$.\label{fig:metro_ridership}}
\end{figure*}

We gathered the relevant data for $138$ metro systems across the world (see Materials and Methods), which we cross-verified when possible with the data given by network operators. We plot the ridership $R$ as a function of $N_s\,\rho$ on Fig.~\ref{fig:metro_ridership} (left) and observe that the data is consistent with a linear behavior. We measure a slope of $800\, \text{km}^2/\text{year}$ which gives an estimate for $d_0$
\begin{equation}
d_0 \approx 500\,\text{m}
\end{equation}
We illustrate this result on Fig.~\ref{fig:metro_ridership} (right) by representing each subway stations of Paris with a circle of radius $500\,\text{m}$.
%

So far, the distance $d_0$ appears here an intrinsic feature of user's behaviors: it is the maximal distance that an individual would
walk to go to a subway station.

\begin{figure*}
\centering
\includegraphics[width=0.49\textwidth]{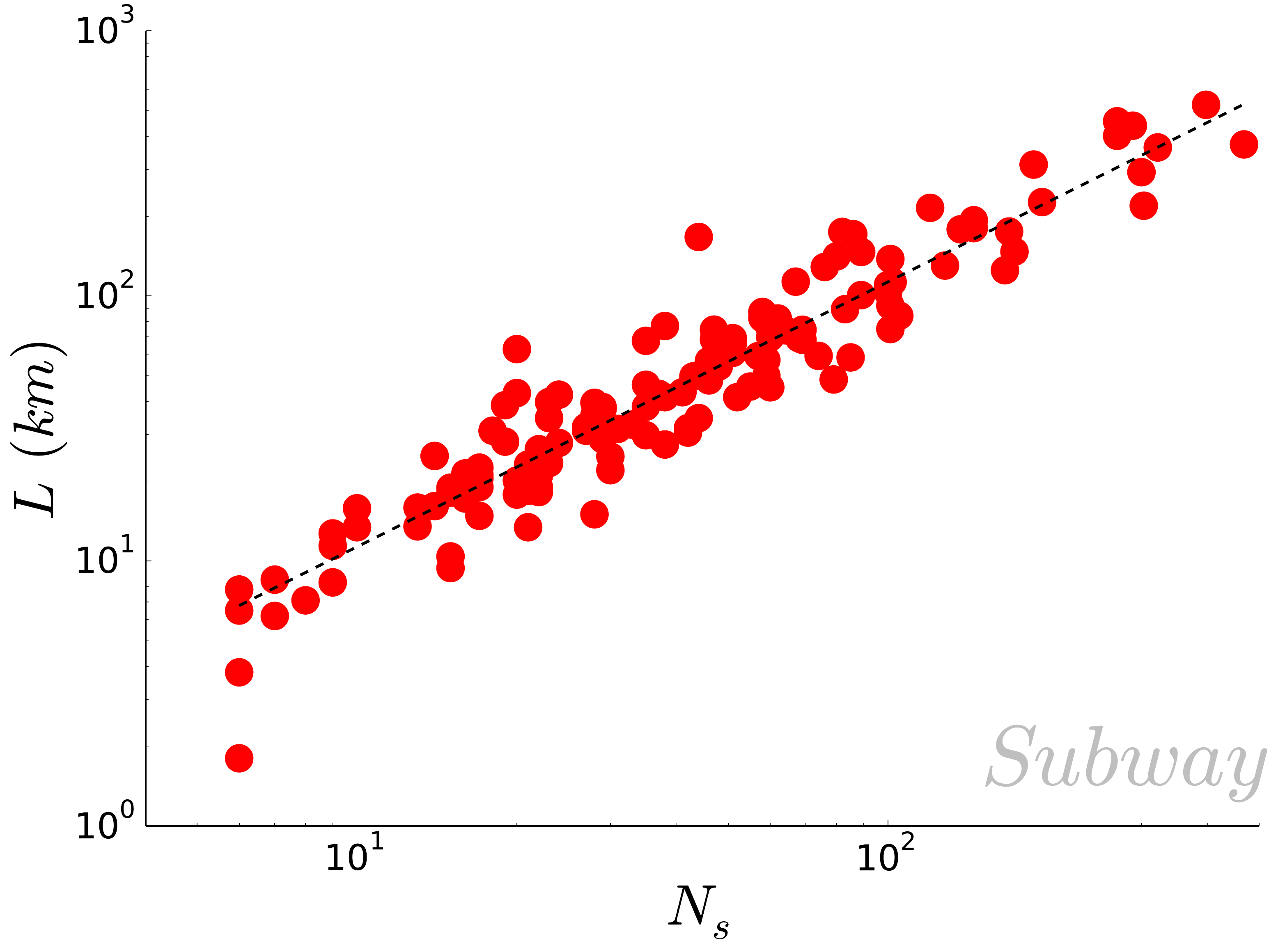}
\includegraphics[width=0.49\textwidth]{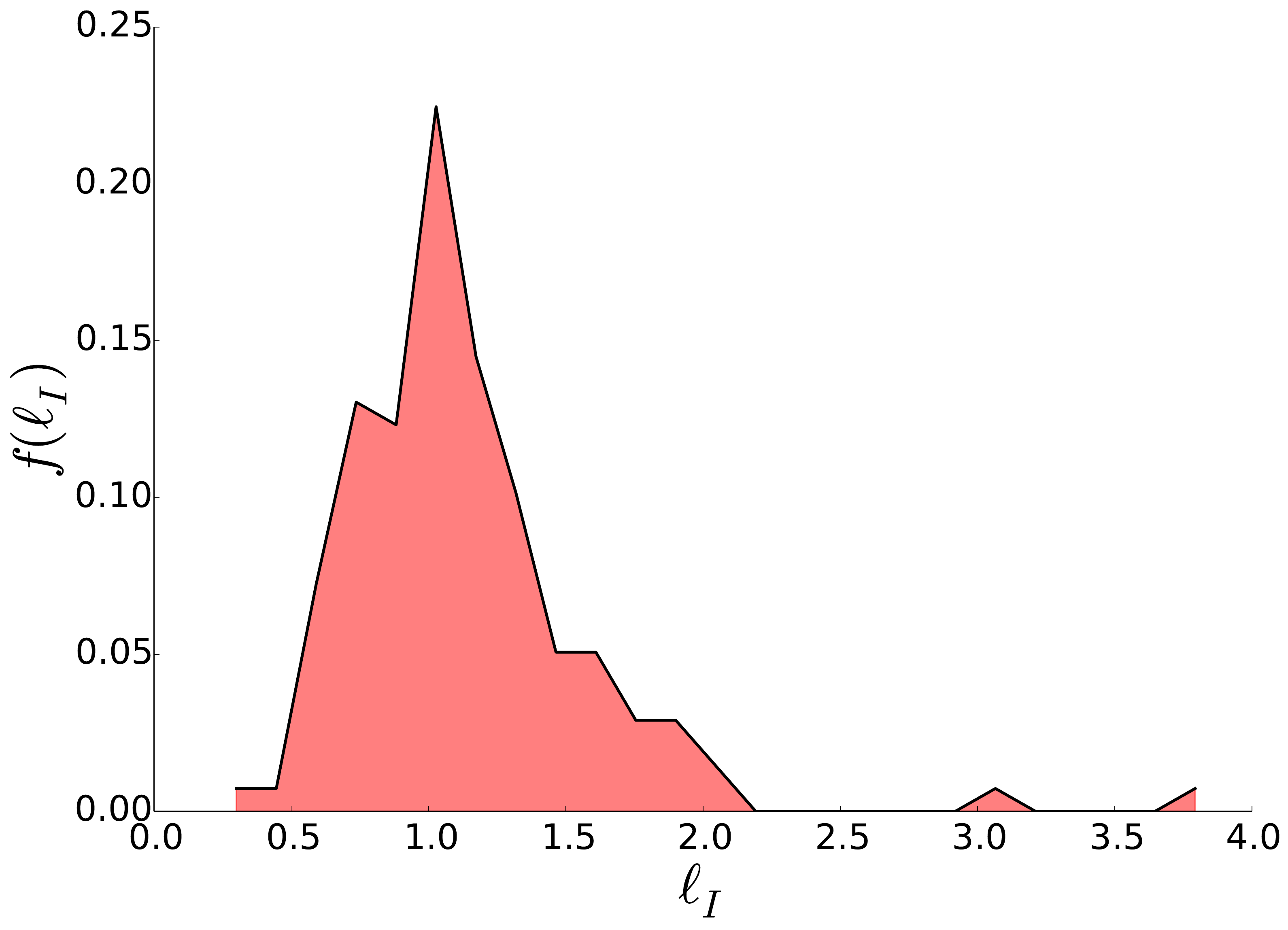}
\caption{{\bf (Subway) Relation between the length and the number of stations} (Left) Length of $138$ subway networks in the world as a function of the number of stations. A linear fit gives $L \sim 1.13\,N_S\,(R^2=0.93)$ (Right) Empirical distribution of the inter-station length. The average interstation distance is found to be $\overline{\ell_1} \approx 1.2\, \text{km}$ and the relative standard deviation is approximately $440\,\text{m}$. \label{fig:metro_length_stations}}
\end{figure*}

The average interstation distance $\ell_1$ is another distance characteristic of the subway system. Rigorously, this distance depends on the average degree $<k>$ of the network so that $\ell_1 = \frac{2\,L}{N_s <k>}$. It has however been found \cite{Roth:2012} that for the $13$ largest subway systems in the world, $<k> \in \left[2.1,2.4\right]$, so that we can reasonably take $<k> / 2 \approx 1$ and thus
\begin{equation}
\ell_1 \simeq \frac{L}{N_s}
\end{equation}
The interstation distance depends in general on many technological and economical parameters, but we expect that for a properly designed system it will match human constraints. Indeed, if $d_0\ll\ell_1$, the network is not dense enough and in the opposite case $d_0\gg\ell_1$, the system is not economically interesting. We can thus reasonably expect that the interstation distance fluctuates slightly around an average value given by twice the typical station attraction distance $d_0$
\begin{equation}
d_0 = \frac{\ell_1}{2} = \frac{L}{2\,N_s}
\end{equation}
It follows from this assumption that the interstation distance is constant and independent from  the population size. In order to test our assumption, we plot on Fig.~\ref{fig:metro_length_stations} (left) the total length of subway networks as a function of the number of stations. The data agrees well with a linear fit $L \sim 1.13\,N_S\,(r^2=0.93)$. We also plot on Fig.~\ref{fig:metro_length_stations} (right) the normalized histogram of the inter-station length, showing that the interstation distance is indeed narrowly distributed around an average value $\overline{\ell_1} \approx 1.2\,\text{km}$ with a variance $\sigma \approx 400\,\text{m}$, consistently with the value found above for $d_0\approx 500\,\text{m}$. The outliers are San Francisco, whose subway system is more of a suburban rail service and Dalian, a very large chinese city whose metro system is very young and still under development.

As a result of the previous argument, we can express $\ell_1$ in terms of the systems characteristics. Indeed, the total ridership now reads

\begin{equation}
R \sim \overline{\xi}\pi\rho\frac{L^2}{N_s}
\label{eq:ridership-other}
\end{equation}
If we assume to be in the steady-state $Z_{sub} \approx 0$, using the results from Eqs.~(\ref{eq:cost-benefit},\ref{eq:ridership-other}), we find that the total length of the network and the number of stations are linked at first order in $\epsilon_s/\epsilon_L$ by
\begin{equation}
L \sim \left( \frac{4 \epsilon_L}{\pi\,\xi\,f\,\rho} + \frac{\epsilon_s}{\epsilon_L}\right) N_s
\label{eq:length-stations}
\end{equation}
and that the interstation distance reads
\begin{equation}
\ell_1 = \frac{4 \epsilon_L}{\pi\,\xi\,f\,\rho} + \frac{\epsilon_s}{\epsilon_L}
\end{equation}
This relation implies that the interstation distance increases with the station maintenance cost, and decreases with increasing line maintenance costs, density and fare. We thus see that the adjustment of $\ell_1$ to match $2\,d_0$ can be made through the fare price (or subsidies by the local authorities or national government). At this point, it would be interesting to get reliable data about the maintenance costs and fares for subway systems in order to pursue in this direction and to test the accuracy of this prediction.

\begin{figure*}
\centering
\includegraphics[width=0.49\textwidth]{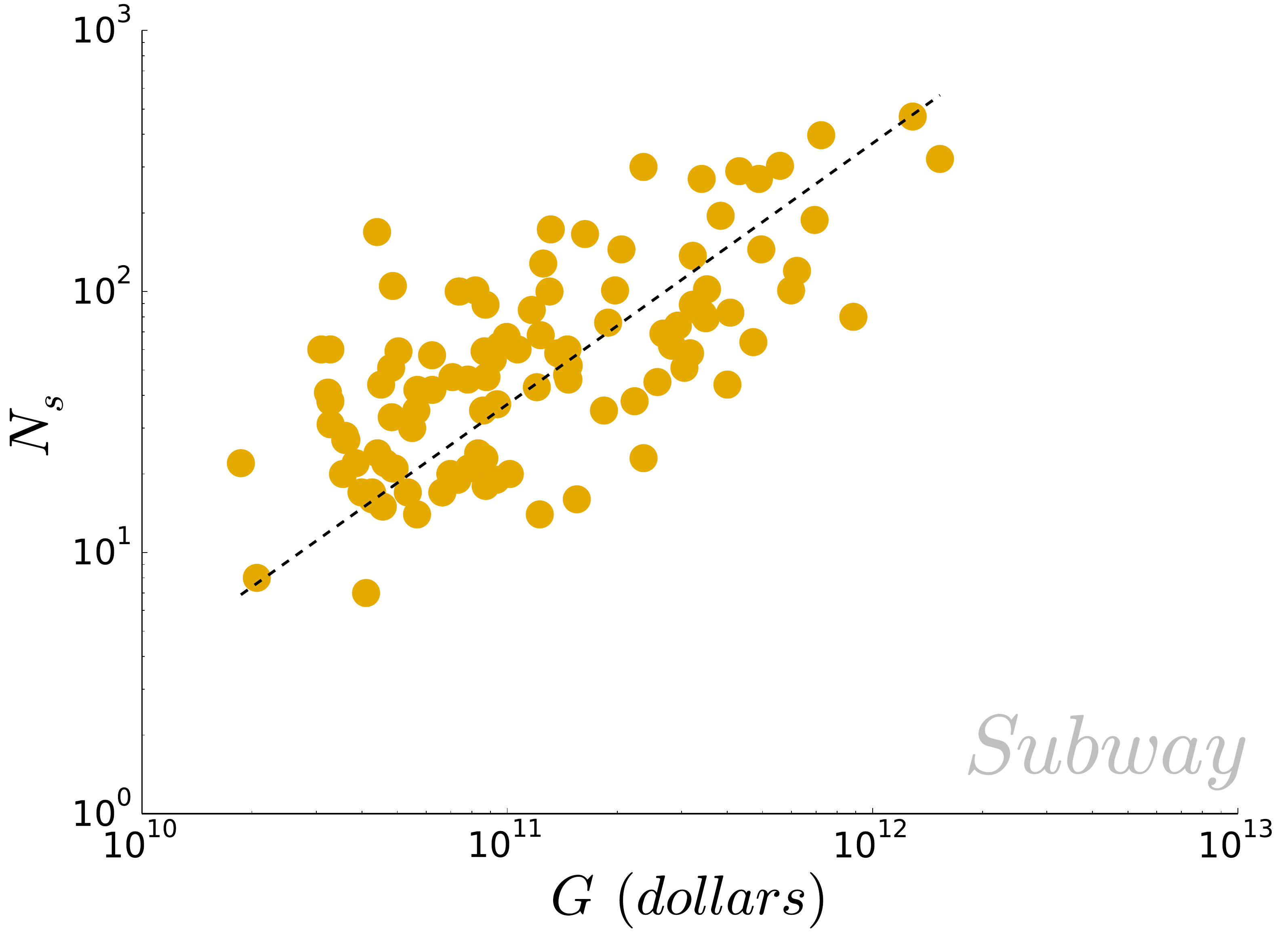}
\includegraphics[width=0.49\textwidth]{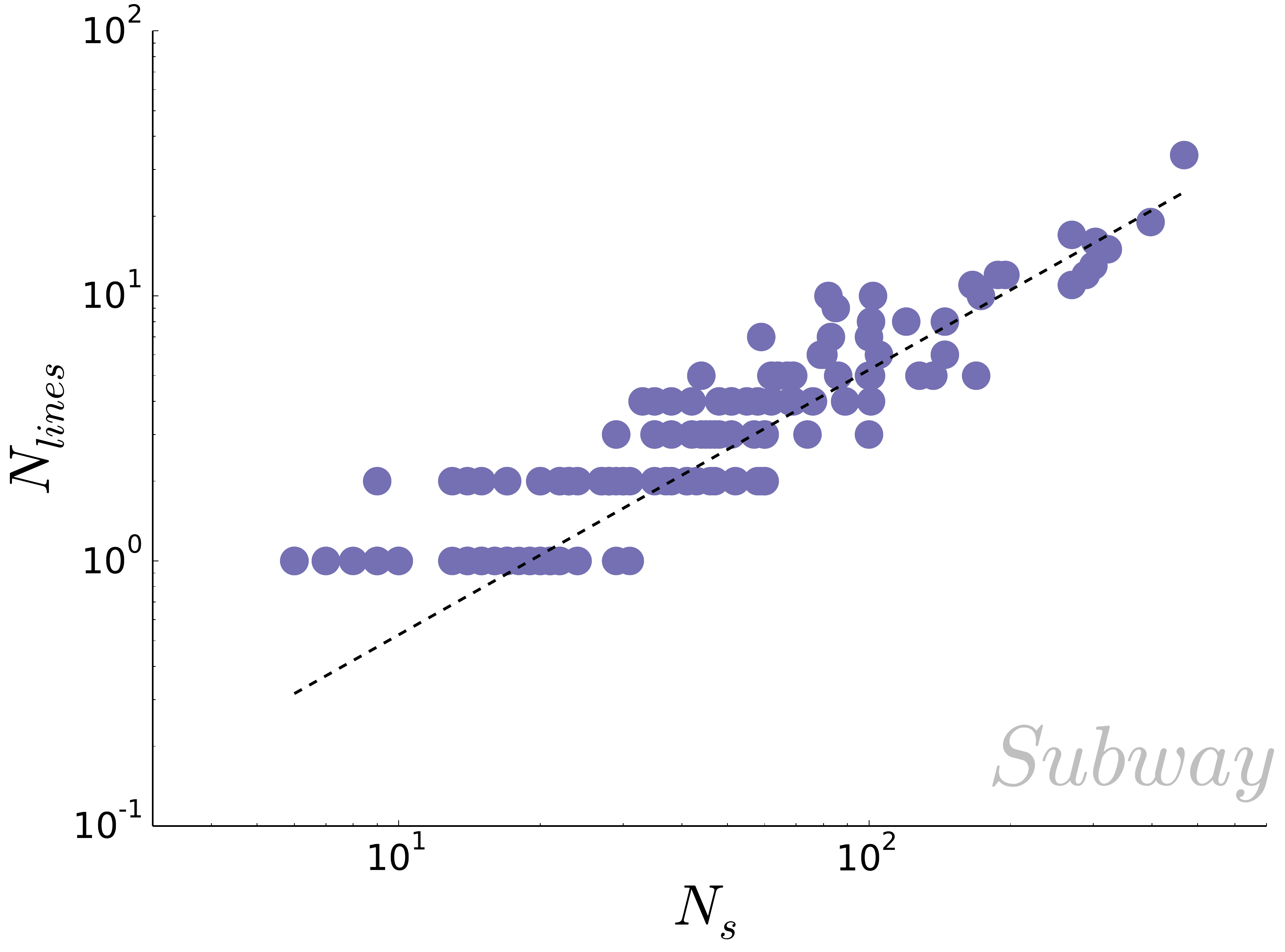}
\caption{{\bf (Subway) Size of the subway system and city's wealth} (Left) We plot the number of stations for the different subway systems in the dataset as a function of the Gross Metropolitan Product of the corresponding cities (obtained for $106$ subway systems). A linear fit (dashed line) gives $N_s = 2.51\, 10^{-10}\,G$ ($R^2=0.73$). {\bf (Subway) Number of lines and number of stations} (Right) We plot the number of metro lines $N_{lines}$ as a function of the number of stations $N_s$. A linear fit on the $138$ data points gives $N_{lines} \approx 0.053\,N_s\,(R^2=0.93)$, or, in other words, metro lines comprise on average $19$ stations.}
\label{fig:metro_stations_gdp}
\end{figure*}

So far, we have a relation between the total length and the number of stations, but we need another equation in order to compute their value. Intuitively, it is clear that the number of stations -- or equivalently the total length -- of a subway system is an increasing function of the wealth of the city. We assume a simple, linear relation of the form
\begin{equation}
N_s = \beta \frac{G}{\epsilon_s}
\label{eq:ns_G}
\end{equation}

where $G$ is the city's Gross Metropolitan Product (GMP), and $\beta$ the fraction of the city's wealth invested in public transportation. This relation can equivalently be interpreted as the proportional relation between the number of station per person and the city's development, as measured by its GMP per capita. On Fig.~\ref{fig:metro_stations_gdp} (left) we plot the number of stations of different metro systems around the world as a function of the Gross Metropolitan Product of the corresponding city. A linear fit agrees relatively well with the data ($R^2=0.73$, dashed line), and gives $\frac{\epsilon_s}{\beta} \approx 10^{10}\,\text{dollars/station}$. However, the dispersion around the linear average behaviour is important: more specific data is needed in order to investigate whether differences in the construction costs and investments (or the age of the system) can explain the dispersion, or if other important parameters need to be taken into account. Incidentally, another possibility would be to assume that the size of the system depends on the age of the system or the development of the city (measured by the GMP per capita). However, in both cases, we found poor correlations. At this stage, we thus conclude that the number of stations (respectively the density of stations) mostly depends on the total GMP (respectively the GMP per capita).

%

Finally, we consider the number of different lines with distinct tracks. A natural question is how the number of lines $N_{lines}$ scales with the number stations $N_s$, that is to say whether lines get proportionally smaller, larger or the same with the size of the whole system. We plot the number of lines as a number of stations on Fig.~\ref{fig:metro_stations_gdp} (right) and find that the data agree with a linear relationship between both quantities ($R^2=0.93$). In other words, the number of stations per line is distributed around a typical value of $19$, whatever the size of the system.

\subsection*{Railway networks}

\begin{figure}
\centering
\includegraphics[width=0.5\textwidth]{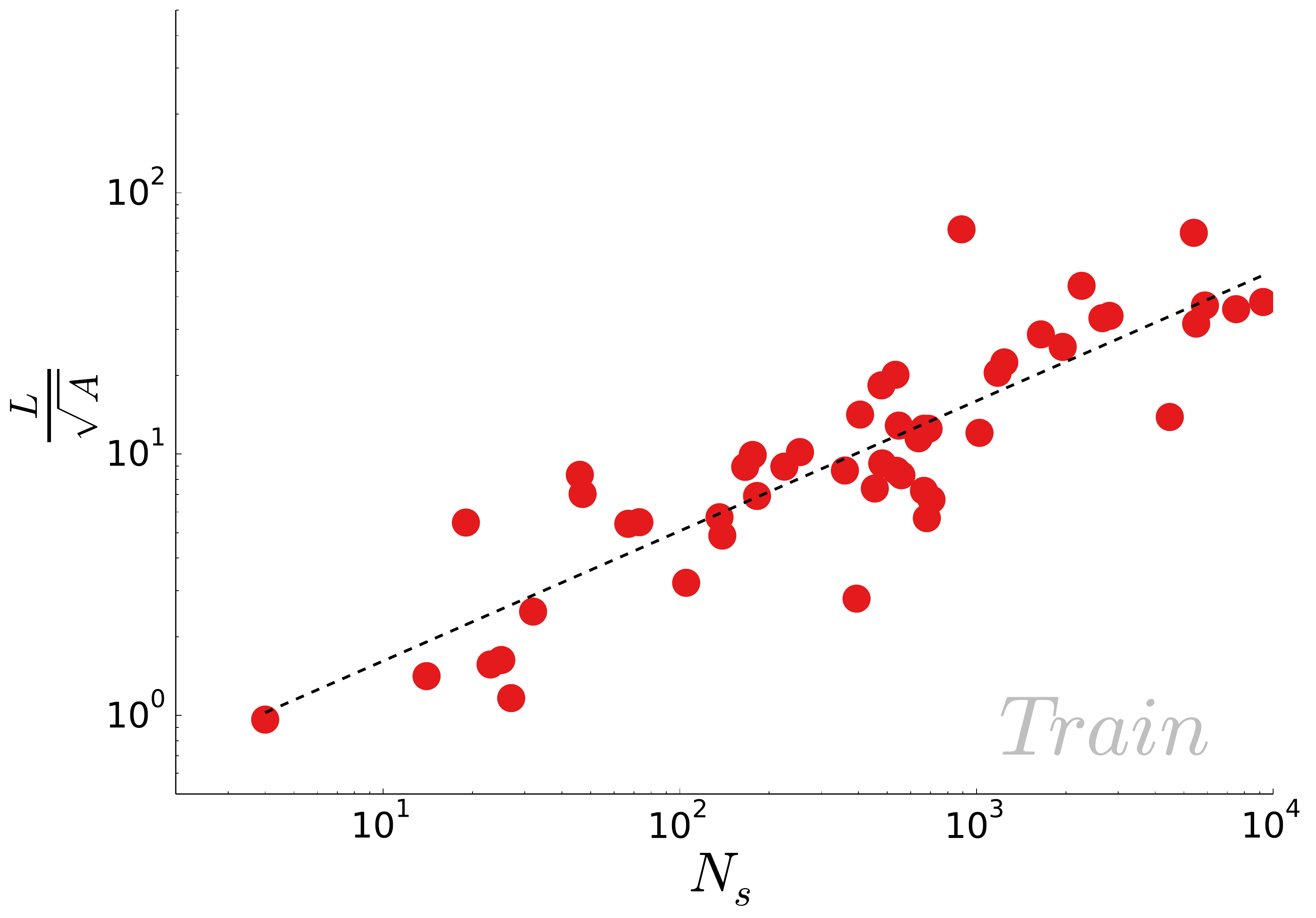}
\caption{{\bf (Train) Total length and number of stations} Total length of the national railway network $L$ rescaled by the typical size of the country $\sqrt{A}$ as a function of the number of stations $N_s$. The dashed line shows the best power-law fit on the $50$ data points with an exponent $0.50 \pm 0.08\,(R^2 = 0.87)$.\label{fig:length-stations}}
\end{figure}

We first discuss an important difference between railway and subway networks. In the subway case, the interstation distance $\ell_1$ is such that it matches human constraints: $\ell_1\sim 2\,d_0$ where $d_0$ is the typical distance that one would walk to reach a subway station. For the railway network, the logic is however different: while subways are built to allow people to move within a dense urban environment, the purpose of building a railway is to connect different cities in a country. In addition, due to the long distance and hence high costs, it seems reasonable to assume that each city is connected to its closest neighbouring city. In this respect, the railway network appears as a planar graph connecting in an economical way, randomly distributed nodes (cities) in the plane. If we assume that a country has an area $A$ and $N_s$ train stations, the typical distance between nearest stations is
\begin{equation}
\ell_N = \sqrt{\frac{A}{N_s}}
\end{equation}
The total length $L \sim N_s\,\ell_N$ is then given by
\begin{equation}
L \sim \sqrt{A\, N_s}
\end{equation}
In order to test this relation for different countries, we plot the adimensional quantity $\frac{L}{\sqrt{A}}$ as a function of the number of stations $N_s$ on Fig.~\ref{fig:length-stations}. A power law fit gives an exponent $0.50 \pm 0.08\,(R^2 = 0.87)$, which is consistent with the previous argument.

\begin{figure}
\centering
\includegraphics[width=0.5\textwidth]{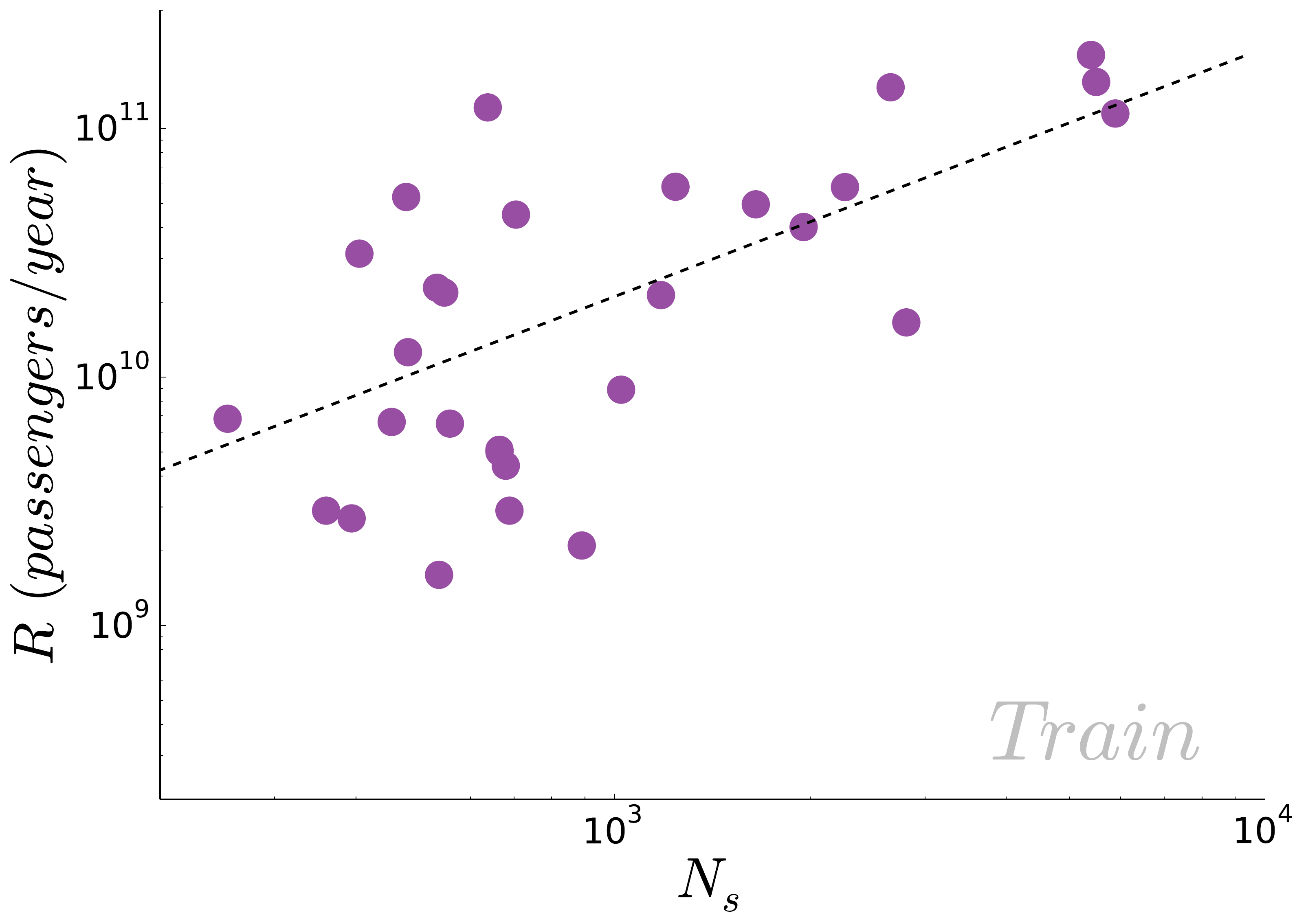}
\caption{{\bf(Train) Ridership and number of stations} The total yearly ridership $R$ of the railway networks as a function of the number of stations. A linear fit on the $47$ data points gives $R \sim 7.0\,10^8\:N_s$ ($R^2 = 0.86$)}
\label{fig:train_rider}
\end{figure}

At this point, we have a relation between $L$ and $N_s$, but we need to find expressions for the other quantities. In contrast with subway systems, due to distances involved, the ticket price usually depends on the distance travelled and we denote by $f_L$ the ticket price per unit distance. The relevant quantity for benefits is therefore not the raw number of passengers -- as in subways -- but rather the total distance travelled on the network $T$. Also, again due to the long distances spanned by the network, the costs of stations can be neglected as a first approximation, and we get for the budget the following expression
\begin{equation}
Z_{train} \simeq T\, f_L - \epsilon_L\, L
\end{equation}
In the steady-state regime $Z_{train}\approx 0$, or in other words the revenue generated by the network use must be of the order of the total maintenance costs~\cite{Louf:2013}, which leads to
\begin{equation}
T \sim \frac{\epsilon_L}{f_L} L
\end{equation}
In addition, if we assume that the order of magnitude of a trip is given by $\ell_N$, the total travelled length is simply proportional to the ridership
$T\sim \ell_N R$ leading to 
\begin{equation}
R \sim \frac{\epsilon_LN_s}{f_L}
\end{equation}
We thus plot the total daily ridership $R$ as a function of the total number of stations $N_s$ (figure \ref{fig:train_rider}), and despite the small number of available data points, a linear relationship between these both quantities seems to agree with empirical data on average ($R^2 = 0.86$). This result should be taken with caution, however, due to the important dispersion that is observed around the average behaviour, and the small number of observations.

\begin{figure}
\centering
\includegraphics[width=0.5\textwidth]{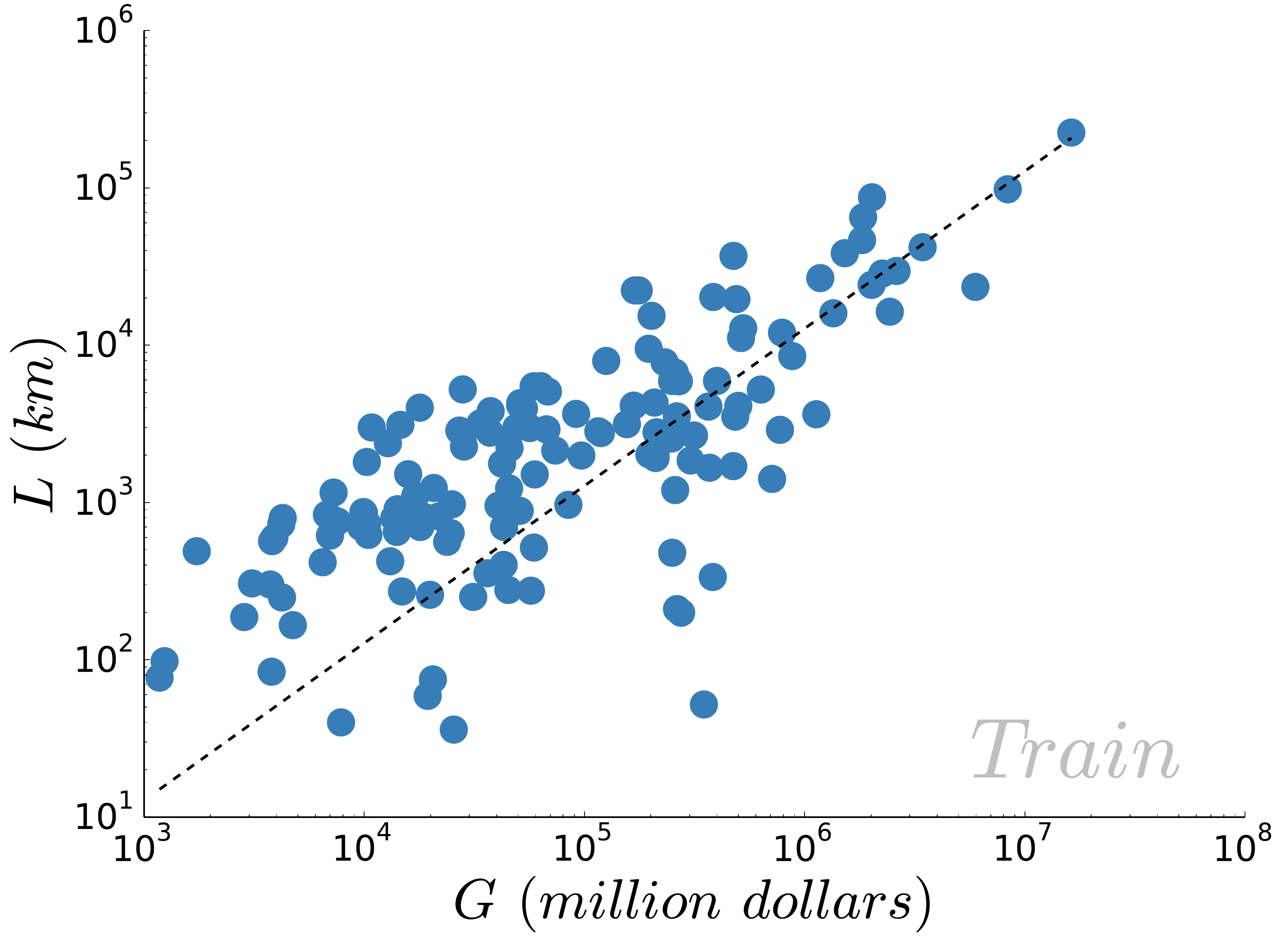}
\caption{{\bf(Train) Total length of the network and wealth} Total length of the railway network $L$ as a function of the country GDP $G$. The dashed line shows the linear fit on the $138$ data points which gives $\epsilon_L / \alpha \approx 10^4\, \text{dollars.km}^{-1}\,(R^2 = 0.91)$.\label{fig:length-gdp}}
\end{figure}

According to the previous result, the total length and the number of stations are related to each other. We now would like to understand what property of the underlying country determines the total length of the network. That is to say, why networks are longer in some countries than in others. As in subway systems, economical reasons seem appealing. Indeed, the railway networks of some large african countries such as Nigeria are way smaller than that of countries such as France or the UK of similar surface areas. A priori, when estimating the cost of a railway network, one should take into account both the costs of building lines and the stations. However, as stated above, considering the distances involved, the cost of building a station is negligible compared to that of building the actual lines. We thus can reasonably expect to have
\begin{equation}
L \sim \frac{\alpha\,G}{\epsilon_L}
\end{equation}
where $G$ is here the country's Gross Domestic Product (GDP) used as an indicator of the country's wealth, and $\alpha < 1$ the ratio of the GDP invested in railway transportation. We plot $L$ as a function of $G$ on Fig.~\ref{fig:length-gdp} and the data agree well ($R^2 = 0.91$) with a linear dependence between $L$ and $G$ (note that we have more points here due to the fact that the data about the total length of a railway network is easier to get). 
Again, the dispersion indicates that the linear trend should only be understood as an average behaviour and that local particularities can have a strong impact on the important deviations observed. For instance, the United Arab Emirates are far from the average behaviour, with a $52\,\text{km}$ network and a GDP of roughly $3\times 10^5$ million dollars. Yet, the construction of a $1,200\,\text{km}$ railway network has been decided in 2010, which would bring the country closer to the average behaviour.
As in the case of subways, we also tried to see whether $L$ could better be explained by the development of the country, as measured by its GDP per capita, but we didn't find any significant correlations.

\section*{Discussion}

We observed scaling relations for global properties of railways and subways and the existence of such relations suggests that basic, common mechanisms are at play during their evolution. A probable reason for the presence of these systems is the mobility demand and their structure is driven by economical mechanisms that seem to be the same for all countries, independently from any cultural, or historical considerations. The fact that macroscopic properties seem to be independent from specific details opens the possibility for simple modelling, and in this spirit, we have proposed a general framework to connect the properties of railway and subway systems (ridership, total length and number of stations) to the socio-economic and spatial characteristics (population, area, GDP) of the country or city where they are built. Despite their simplicity, our arguments agree satisfactorily with data we gathered for almost $140$ subway systems and $50$ railway networks accross the world. As a result, and maybe surprisingly, the knowledge of simple characteristics of a country or a city are enough to give an estimate of the size and use of its transportation system.

It should be noted that the noise associated with the data (and sometimes their definition, see Material and Methods) makes it difficult to infer behaviours from the empirical analysis alone. Therefore, the most appropriate way to proceed, we believe, is to make assumptions about the systems and build a model whose predictions can then be tested against data.

This study suggests that the fundamental difference between railways and subways comes from the determination of the interstation distance. While it is imposed by human constraints in the subway case, the railway network has to adapt to the spatial distribution of cities in a country. This remark is at the heart of the different behaviors observed for railways and subways (see Table~\ref{table:summary} for a summary of these differences).

The previous arguments are able to explain the average behaviour of various quantities. Nevertheless, it would be interesting to identify deviations from these behaviours, and see as suggested in \cite{Derrible:2009} whether they are correlated with topological properties of the system, or other properties of the network and the region. We think that the relations presented here provide however a simple framework within which local particularities can be discussed and understood. We also think that this framework could serve as a useful null-model to quantify the efficiency of individual transportation networks, and compare them to each other. This would however require more specific data than those that were available to us. 

While we have focused on an average, static description of metro systems, we believe that our study provides a better understanding of how these systems interact with the region they serve. This new insight is a necessary step towards a model for the growth of subway systems that takes the characteristics of the city into account. Indeed, although models of network growth exist, the length of networks and nodes at a given time is usually imposed exogeneously, instead of being linked to the socio-economic properties of the substrate. This study provides a simple approach to these complex problems and could help in building more realistic models, with less exogeneous parameters.

It would also be interesting to gather data about the exact structure of all the networks, to study whether there is a relationship between their topology (degree distribution, detour index, etc.) and properties of the substrate, as was done for the road network in~\cite{Levinson:2012}.

Finally, gathering historical data should allow to address the problem of the conditions for the appearance of a subway in a city. Indeed, we observe empirically that the GDP of the cities that have a subway system is always larger than about $10^{10}$ dollars, a fact that calls for a theoretical explanation. 

\begin{table}[!ht]
\centering
\caption{
\bf{Summary of the differences between subways and railways} \label{table:summary}}
\begin{tabular}{|c|c|c|}
\hline
 & {\bf Subway} & {\bf Train} \\
 \hline
$L / N_s$ & cste. & $\sqrt{\frac{A}{N_s}}$\\
$R$ & $\frac{P}{A}\,N_s$ & $N_s$ \\
$G$ & $N_s$ & $L$ \\
\hline
\end{tabular}
\begin{flushleft} We summarize the difference of behaviour between subways and railways. The scaling of the length $L$ of the network with the number of stations $N_s$ reveals the different logics behind the growth of these systems. Another difference lies in the total ridership $R$: while it depends also on the population density $P/A$ for subways, it only depends on the number of stations $N_s$ for train networks. Finally, the size of both types of networks can be expressed as a function of the wealth of the region, represented here by the GDP $G$. However, because the interstation length is constant for subways, the size is expressed in terms of the number of stations $N_s$; in the case of railway networks, the cost of stations is negligible compared to the building cost of lines, and the size is expressed in terms of the total length $L$.
\end{flushleft}
\label{tab:label}
\end{table}


\section*{Materials and Methods}

Data for $138$ subways accross the world were mainly collected on Wikipedia~\cite{Wiki}, and cross-referenced with the operators' data when possible. The cities' GDP per capita was retrieved for $114$ cities from Brooking's Global MetroMonitor~\cite{Brookings}. The choice of population and city area was more subtle. Indeed, most subway systems span an area greater than the city core, and the relevant area therefore lies somewhere between the city core's area and the total urbanized area. We chose to use the population and surface area data for urbanized areas provided by Demographia~\cite{Demographia}.

While data about ridership, network length were easily retrievable for more than $100$ countries from the UIC Railisa 2011 database~\cite{railisa}, data about the number of stations were more difficult to find. We had to use various data sources, mainly scrapping the operators' ticket booking websites. Data about the GDP, population and surface areas of different countries were obtained from the World Bank~\cite{WorldBank}, and the United Nations Statistics Division~\cite{UnitedNations}.\\

All the data used for this study are publicly available in tsv format at~\cite{data_repo}.

\section*{Acknowledgments}

We thank Giulia Ajmone-Marsan and R\"udiger Ahrend for interesting discussions at an early stage of this project.


\clearpage


\begin{thebibliography}{99}

\bibitem{Benguigui:1992} Benguigui L (1992) The fractal dimension of some railway networks. Journal de Physique I 2:
385--388.

\bibitem{Benguigui:1995} Benguigui L (1995) A fractal analysis of the public transportation system of paris. Environment
and Planning A 27: 1147--1161.

\bibitem{Derrible:2009} Derrible S, Kennedy C (2009) Network analysis of world subway systems using updated graph
theory. Transportation Research Record: Journal of the Transportation Research Board 2112: 17--25.

\bibitem{Sienkiewicz:2005} Sienkiewicz J, Holyst JA (2005) Statistical analysis of 22 public transport networks in poland.
Physical Review E 72: 046127.

\bibitem{Levinson:2012} Levinson D (2012) Network structure and city size. PLoS ONE 7: e29721.

\bibitem{VonFerbe:2009} Von Ferber C, Holovatch T, Holovatch Y, Palchykov V (2009) Modeling metropolis public transport. In: Traffic and Granular Flow, Springer. pp. 709–719.

\bibitem{Roth:2012} Roth C, Kang SM, Batty M, Barthelemy M (2012) A long-time limit for world subway networks. Journal of The Royal Society Interface 9: 25402550.

\bibitem{Leng:2014} Leng B, Zhao X, Xiong Z (2014) Evaluating the evolution of subway networks: Evidence from beijing subway network. EPL (Europhysics Letters) 105: 58004.

\bibitem{Levinson:2008} Levinson D (2008) Density and dispersion: the co-development of land use and rail in london. Journal of Economic Geography 8: 5577.

\bibitem{Xie:2009} Xie F, Levinson D (2009) Topological evolution of surface transportation networks. Computers, Environment and Urban Systems 33: 211--223.

\bibitem{Louf:2013} Louf R, Jensen P, Barthelemy M (2013) Emergence of hierarchy in cost-driven growth of spatial networks. Proceedings of the National Academy of Sciences 110: 8824--8829.

\bibitem{Kansky:1963} Kansky KJ (1963) Structure of transportation networks: relationships between network geometry and regional characteristics. PhD Thesis.

\bibitem{Banavar:1999} Banavar JR, Maritan A, Rinaldo A (1999) Size and form in efficient transportation networks. Nature 399: 130--132.

\bibitem{Louf:2014} Louf R, Barthelemy M (2014) How congestion shapes cities: from mobility patterns to scaling. Sci. Rep. 4: 5561 (2014).

\bibitem{Black:1971} Black WR (1971) An iterative model for generating transportation networks. Geographical Analysis 3: 283288.

\bibitem{Matsunaka:2013} Matsunaka R, Oba T, Nakagawa D, Nagao M, Nawrocki J (2013) International comparison of the relationship between urban structure and the service level of urban public transportation: A comprehensive analysis in local cities in japan, france and germany. Transport Policy 30: 26--39.

\bibitem{Kuby:2004} Kuby M, Barranda A, Upchurch C (2004) Factors influencing light-rail station boardings in the united states. Transportation Research Part A: Policy and Practice 38: 223--247.

\bibitem{Wiki} Data about subway length, number of stations are available on the wikipedia website \url{http://www.wikipedia.org}.

\bibitem{Brookings} GMP per capita data for different cities accross the world are available on Brooking’s Global Metromonitor website \url{http://www.brookings.edu/research/interactives/global-metro-monitor-3}.

\bibitem{Demographia} Surface area and population data for urbanized areas accross the world are available on the Demographia website \url{http://www.demographia.com}.

\bibitem{railisa} The Railisa database is available on the UIC’s website \url{http://www.uic.org/spip.php?article1353}.

\bibitem{WorldBank} Data about the gdp of countries were taken from the World Bank’s website \url{http://data.worldbank.org/indicator/NY.GDP.MKTP.CD}.

\bibitem{UnitedNations} Population and surface areas of countries were taken from the Demographic Yearbook, available on the United Nation Statistical Division’s website \url{http://unstats.un.org/unsd/demographic/products/dyb/dyb2.htm}.

\bibitem{data_repo} All the data used for this study are available in tsv format at \url{http://github.com/rlouf/data/tree/master/scaling_transportation}.
\end{thebibliography}
\end{document}